\documentclass[twocolumn]{aastex631}

\let\tablenum\relax
\usepackage{hyperref}

\usepackage[separate-uncertainty,multi-part-units = single,group-digits=false]{siunitx}
\usepackage{bm}

\usepackage{chemformula}
\usepackage{amsmath}
\usepackage{float}

\setcitestyle{notesep={ }}
\shorttitle{Dynamical Mass of PZ Tel B}
\shortauthors{Franson \& Bowler}

\graphicspath{{./}{figures/}}

\begin{document}

\title{Dynamical Mass of the Young Brown Dwarf Companion PZ Tel B}

\author[0000-0003-4557-414X]{Kyle Franson}
\altaffiliation{NSF Graduate Research Fellow} \affiliation{Department of Astronomy, The University of Texas at Austin, Austin, TX 78712, USA}

\author[0000-0003-2649-2288]{Brendan P. Bowler}
\affiliation{Department of Astronomy, The University of Texas at Austin, Austin, TX 78712, USA}

\begin{abstract}
    Dynamical masses of giant planets and brown dwarfs are critical tools for empirically validating substellar evolutionary models and their underlying assumptions. We present a measurement of the dynamical mass and an updated orbit of PZ Tel B, a young brown dwarf companion orbiting a late-G member of the $\beta$ Pic moving group. PZ Tel A exhibits an astrometric acceleration between Hipparcos and Gaia EDR3, which enables the direct determination of the companion's mass. We have also acquired new Keck/NIRC2 adaptive optics imaging of the system, which increases the total baseline of relative astrometry to 15 years. Our joint orbit fit yields a dynamical mass of $27^{+25}_{-9} \, M_{\mathrm{Jup}}$, semi-major axis of $27^{+14}_{-4} \, \mathrm{au}$, eccentricity of $0.52^{+0.08}_{-0.10}$, and inclination of $91.73^{+0.36}_{-0.32} {}^\circ$. The companion's mass is consistent within $1.1\sigma$ of predictions from four grids of hot-start evolutionary models. The joint orbit fit also indicates a more modest eccentricity of PZ Tel B than previous results. PZ Tel joins a small number of young (${<}200 \, \mathrm{Myr}$) systems with benchmark substellar companions that have dynamical masses and precise ages from moving group membership.
\end{abstract}

\keywords{Brown dwarfs (185) --- Direct imaging (387) --- Astrometry (80) --- Orbit determination (1175)}

\section{Introduction \label{sec:intro}}
Model-independent dynamical masses of giant planets and brown dwarfs offer unique probes into the formation and evolution of substellar objects. When paired with a bolometric luminosity and age constraint, these objects facilitate precision tests of substellar cooling models and their underlying assumptions about the initial entropy, atmospheric physics, and internal structure of low-temperature objects. These benchmark systems are rare, largely owing to the low intrinsic rate of substellar companions within a few tens to hundreds of au. The Gemini Planet Imager Exoplanet Survey measured an occurrence rate of $0.8^{+0.8}_{-0.5}\%$ for brown dwarf companions ($13{-}80 \, M_\mathrm{Jup}$) around all stars ($0.2{-}5 \, M_\odot$) at semi-major axes from $10{-}100 \, \mathrm{au}$. \citep{nielsenGeminiPlanetImager_2019}. The SpHere INfrared Exoplanet survey measured a slightly higher occurrence rate of $5.8^{+4.7}_{-2.8} \%$ for substellar companions ($1{-}75 \, M_\mathrm{Jup}$) from $5{-}300 \, \mathrm{au}$ around FGK stars \citep{viganSphereInfraredSurvey_2021}. These are comparable to earlier studies that found brown dwarf companion frequencies of $\approx$1--4\% across all separations (e.g., \citealt{metchevPalomarKeckAdaptive_2009}; \citealt{brandtStatisticalAnalysisSeeds_2014}; \citealt{lafreniereAdaptiveOpticsMultiplicity_2014}; see \citealt{bowlerOccurrenceRatesDirect_2018} and references therein).

The sample of substellar companions with dynamical masses and age constraints now amounts to ${\approx} 20$ brown dwarf companions \citep[e.g.,][]{bowlerOrbitDynamicalMass_2018,cheethamDirectImagingUltracool_2018,brandtPreciseDynamicalMasses_2019,brandtDynamicalMass70_2020,brandtImprovedDynamicalMasses_2021,kuzuharaDirectImagingDiscovery_2022,bonavitaResultsCopainsPilot_2022,fransonDynamicalMassYoung_2022,fransonAstrometricAccelerationsDynamical_2023,liSurveyingNearbyBrown_2023} and giant planets \citep{dupuyModelindependentMassModerate_2019,lagrangeEvidenceAdditionalPlanet_2019,nowakDirectConfirmationRadialvelocity_2020,brandtPreciseDynamicalMasses_2021,hinkleyDirectDiscoveryInner_2022,mesaAfLepLowest_2023,derosaDirectImagingDiscovery_2023,fransonAstrometricAccelerationsDynamical_2023a}. Most of these benchmark companions orbit old field stars (${>}1\, \mathrm{Gyr}$); only three planets---$\beta$ Pic b, $\beta$ Pic c, and AF Lep b---are young (${<}200 \, \mathrm{Myr}$) and have precise ages based on membership in kinematic associations \citep[][]{dupuyModelindependentMassModerate_2019,nowakDirectConfirmationRadialvelocity_2020,brandtPreciseDynamicalMasses_2021,fransonAstrometricAccelerationsDynamical_2023a,derosaDirectImagingDiscovery_2023,mesaAfLepLowest_2023}.\footnote{The dynamical mass of HR 8799 e has been measured \citep{brandtFirstDynamicalMass_2021}, but the age of the host star is not precisely determined. \citet{zuckermanTucanaHorologiumColumba_2011} and \citet{leeDevelopmentModelsNearby_2019} have linked the system to the Columba and $\beta$ Pic moving groups, while \citet{faramazDetailedCharacterizationHr_2021} later found that HR 8799 is likely a young field star. The ages of the Columba and $\beta$ Pic moving groups are $45^{+11}_{-7} \, \mathrm{Myr}$ and $24 \pm 3 \, \mathrm{Myr}$, respectively \citep{bellSelfconsistentAbsoluteIsochronal_2015}. \citet{sepulvedaDynamicalMassExoplanet_2022} derived an independent age constraint of $10{-}23 \, \mathrm{Myr}$ through fitting isochrones to the mass and fundamental properties of the host star.} Age-dating individual field stars is notoriously challenging, which can limit the precision of model tests for benchmark companions around young field stars \citep[see e.g., HD 984 B;][]{fransonDynamicalMassYoung_2022} compared to objects in young moving groups.

The $\beta$ Pictoris moving group \citep[][]{barradoynavascuesAgePictoris_1999,zuckermanPictorisMovingGroup_2001} is one of the best-studied nearby young kinematic associations. With around 150 bona-fide members and hundreds of additional candidates \citep[e.g.,][]{moorUnveilingNewMembers_2013,shkolnikAllskyComovingRecovery_2017,gagneBanyanXiBanyan_2018}, this association has served as a laboratory for studying young debris disks \citep[e.g.,][]{kalasAsymmetriesBetaPictoris_1995,kalasDiscoveryLargeDust_2004,moorBigSiblingAu_2020,hinkleyDiscoveryEdgeonCircumstellar_2021}, giant planets \citep[e.g.,][]{lagrangeGiantPlanetImaged_2010,liuExtremelyRedYoung_2013,macintoshDiscoverySpectroscopyYoung_2015,dupuyHawaiiInfraredParallax_2018,plavchanPlanetDebrisDisk_2020,nowakDirectConfirmationRadialvelocity_2020,fransonAstrometricAccelerationsDynamical_2023a}, and brown dwarfs \citep[e.g.,][]{billerGeminiNiciPlanetfinding_2010,mugrauerDirectDetectionSubstellar_2010,schneider2massAllwiseSearch_2017,phillips2massJ044356863723033_2020}. The group spans a distance of ${\sim} 20{-}50\, \mathrm{pc}$ \citep{gagneBanyanXiiiFirst_2018} and has an age of \SI{24 \pm 3}{Myr} (\citealt{bellSelfconsistentAbsoluteIsochronal_2015}; see also Table 6 of \citealt{miret-roigDynamicalTracebackAge_2020} for a compilation of age estimates).

PZ Telescopii (${=}$PZ Tel, HD 174429, HIP 92680) is a bright ($V {=} \SI{8.3}{mag}$; \citealt{kiragaAsasPhotometryRosat_2012}) G9IV \citep{torresSearchAssociationsContaining_2006} star at a distance of $47.25 \pm 0.05 \, \mathrm{pc}$ \citep{gaiacollaborationGaiaDataRelease_2022}. It is a well-established member of the $\beta$ Pic moving group \citep{zuckermanPictorisMovingGroup_2001,shkolnikAllskyComovingRecovery_2017}, with BANYAN $\Sigma$ \citep{gagneBanyanXiBanyan_2018} giving a 96.8\% probability of membership based on its Gaia DR3 astrometry and radial velocity measurement of \SI{-3.6 \pm 1.6}{km.s^{-1}} \citep{gaiacollaborationGaiaDataRelease_2022}. A brown dwarf orbiting PZ Tel was independently discovered by \citet{billerGeminiNiciPlanetfinding_2010} and \citet{mugrauerDirectDetectionSubstellar_2010} with Near-Infrared Coronagraphic Imager (NICI; Gemini South) and NaCo (VLT) adaptive optics (AO) imaging, respectively, with a contrast of $\Delta H = 5.38 \pm 0.09 \, \mathrm{mag}$. At the time of its discovery, PZ Tel B was located at a projected separation of ${\sim}350 \, \mathrm{mas}$ (${\sim}16.5\, \mathrm{au}$; \citealt{billerGeminiNiciPlanetfinding_2010,mugrauerDirectDetectionSubstellar_2010}), but has gradually moved outward to over $500 \, \mathrm{mas}$ based on the latest reported astrometric measurement in \citet{stolkerMiraclesAtmosphericCharacterization_2020}. Model-inferred masses in the literature range from about $8 \, M_{\mathrm{Jup}}$ to $60 \, M_\mathrm{Jup}$, but typically fall between $20{-}50 \, M_{\mathrm{Jup}}$ (e.g., $36 \pm 6 \, M_\mathrm{Jup}$, \citealt{billerGeminiNiciPlanetfinding_2010}; $24{-}40 \, M_{\mathrm{Jup}}$, \citealt{mugrauerNewObservationsPz_2012}; $3.3{-}24\, M_{\mathrm{Jup}}$, \citealt{schmidtFirstSpectroscopicObservations_2014}; $45^{+9}_{-7}\, M_{\mathrm{Jup}}$ and $59^{+13}_{-8}\, M_{\mathrm{Jup}}$, \citealt{maireFirstLightVlt_2016}). PZ Tel B has been imaged from 2007 to 2018 with NaCo (VLT), NICI (Gemini South), SINFONI (VLT), and SPHERE (VLT), and shows orbital motion indicating a highly eccentric ($e \gtrsim 0.6$) and nearly edge-on orbit \citep[e.g.,][]{mugrauerNewObservationsPz_2012,ginskiAstrometricFollowupObservations_2014,maireFirstLightVlt_2016,bowlerPopulationlevelEccentricityDistributions_2020}. PZ Tel A exhibits a significant\footnote{$\chi^2 = 21.3$, which corresponds to $4.2\sigma$ for 2 degrees of freedom.} astrometric acceleration of $\SI{3.7 \pm 0.8}{m.s^{-1}yr^{-1}}$ between Hipparcos and Gaia EDR3 in the Hipparcos-Gaia Catalog of Accelerations \citep[HGCA;][]{brandtHipparcosgaiaCatalogAccelerations_2021}, which for the first time enables the measurement of the companion's dynamical mass based on the astrometric reflex motion of the host star.

Here, we perform a joint orbit fit of PZ Tel B incorporating all published astrometry, radial velocities, and the HGCA astrometric acceleration. We also report new Keck/NIRC2 $K_s$-band imaging that expands the total baseline of astrometric observations to 15 years. Through our orbit fit, we determine the dynamical mass of PZ Tel B for the first time and compare it against the predictions of four hot-start substellar evolutionary models using the moving group age and bolometric luminosity of the companion. We conclude by examining the impact of additional epochs of astrometry on the dynamical mass precision.

\begin{figure}
    \centering
    \includegraphics[width=0.9\linewidth]{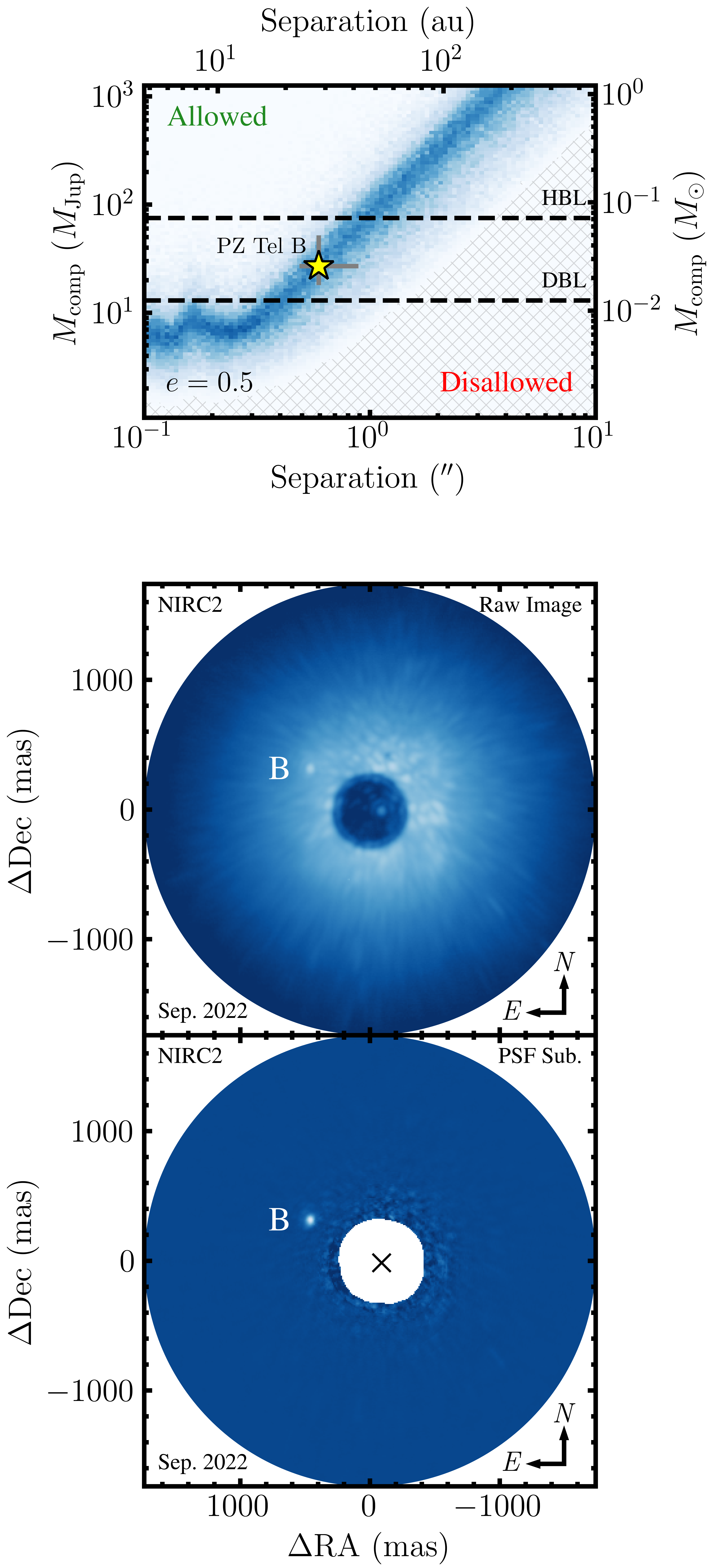}
    \caption{\emph{Top:} Predicted mass of PZ Tel B as a function of projected separation based on its Hipparcos-Gaia astrometric acceleration. These predictions are generated following the procedure detailed in \citet{fransonAstrometricAccelerationsDynamical_2023} except using an eccentricity of 0.5 for our synthetic orbits. The yellow star denotes the companion's dynamical mass ($27^{+25}_{-9} \, M_{\mathrm{Jup}}$) and semi-major axis ($27^{+14}_{-4} \, \mathrm{au}$). \emph{Bottom:} Keck/NIRC2 $K_s$-band coronagraphic imaging of PZ Tel at an airmass of 2.95. The host star is visible behind the partially transparent \SI{600}{mas}-diameter coronagraph. The top frame shows an example raw image from the sequence, while the bottom frame shows a PSF-subtracted image. The images are aligned such that north is up and east is to the left.
    \label{fig:nirc2_image}}
\end{figure}

\section{Keck/NIRC2 Adaptive Optics Imaging \label{sec:obs}}
We observed the PZ Tel system with the NIRC2 camera at W.M. Keck Observatory on UT 2022 September 17. Images were taken in $K_s$ band in field-tracking mode with the \SI{600}{mas} Lyot coronagraph and natural guide star AO \citep{wizinowichAstronomicalScienceAdaptive_2013}. We obtained a total of 10 frames, each consisting of a single coadd with an integration time of \SI{10}{s}. Due to the southern declination of PZ Tel ($\delta \sim -50^\circ$), the observations were taken at an airmass of 2.95 (altitude of $19\fdg6$ at the time of the observation). The median FWHM of the companion PSFs in the sequence is \SI{60}{mas}, which is similar to the Keck $K_s$ diffraction limit of \SI{54}{mas}.

After subtracting darks and flat-fielding the science frames, we remove cosmic rays and bad pixels using the \texttt{L.A.Cosmic} algorithm \citep{vandokkumCosmicRayRejection_2001}. We then correct for geometric distortions from the optics of the imaging system using the solution of \citet{serviceNewDistortionSolution_2016} for the narrow-field mode of the NIRC2 camera. An example frame from the sequence is shown in Figure \ref{fig:nirc2_image}. The companion is visible in the raw coronagraphic frames without the aid of PSF subtraction, while the host star is visible behind the partially transparent coronagraph. We also show a PSF-subtracted image generated by applying the Locally Optimized Combination of Images (LOCI; \citealt{lafreniereNewAlgorithmPointspread_2007}) algorithm with a NIRC2 reference PSF library comprising ${>}2000$ registered coronagraphic frames. Since our observations were not taken with Angular Differential Imaging \citep{maroisAngularDifferentialImaging_2006}, or pupil-tracking mode, we use Reference Star Differential Imaging \citep{lafreniereHstNicmosDetection_2009} with 100 frames selected from the reference library following \citet{bowlerPlanetsLowmassStars_2015a}. We adopt $N_A = 300$, $N_\delta = 1.5$, $g = 1$, and $dr = 2$ for the LOCI parameters. The companion is masked in the reduction.

To measure the astrometry of PZ Tel B, we determine the centroid of the companion and the host star in each raw image, measure the separation ($\rho$) and position angle ($\theta$), and then adopt the mean of those quantities from all frames. The separation values incorporate the NIRC2 plate scale of \SI{9.971 \pm 0.004}{mas/pixel} \citep{serviceNewDistortionSolution_2016}. The position angle is determined from the raw images using the following equation for NIRC2 observations in fixed pupil tracking mode, which uses FITS header values to determine the orientation of celestial north: 
\begin{align}
    \theta = \theta_{\mathrm{raw}} + \mathtt{ROTPOSN} - \mathtt{INSTANGL} - \theta_{\mathrm{north}}.
\end{align}
Here, $\theta_{\mathrm{raw}}$ is the position angle measured from a given frame before north alignment, $\mathtt{ROTPOSN}$ is the instrument rotator position, $\mathtt{INSTANGL}$ is the NIRC2 instrumental position angle zero point of $0\fdg7$, and $\theta_{\mathrm{north}}$ is the angle required to align the detector columns to celestial north ($0\fdg262 \pm 0\fdg020$; \citealt{serviceNewDistortionSolution_2016}).
Following \citet{fransonDynamicalMassYoung_2022}, uncertainties are determined by adding in quadrature the standard deviation of the individual $\rho$ and $\theta$ measurements, the error on the \citet{serviceNewDistortionSolution_2016} distortion correction of $\sigma_d = \SI{1}{mas}$, the plate scale uncertainty, and the uncertainty of the north alignment.

Due to the high airmass of 2.95 at the time of the observation, we correct our astrometry for differential atmospheric refraction (DAR). The impact of DAR on relative astrometry can be divided into monochromatic and chromatic contributions. The monochromatic effect originates from the difference in zenith angle between two sources and has the effect of compressing images along the zenith direction. The chromatic effect occurs due to different wavelengths of light experiencing different levels of refraction in the atmosphere. This causes astrophysical sources to appear to elongate along the zenith direction. Differences in spectral slope between two sources can then alter their respective positions on the detector, influencing the astrometry measured from the ground. 

For each image in the sequence, we compute the monochromatic DAR using Equation 12 of \citet{gublerDifferentialAtmosphericRefraction_1998}. The \texttt{refco} function from the Starlink AST C library \citep{berryAstLibraryModelling_2016} is used to calculate the refraction constants $A$ and $B$ given the relative humidity, temperature, and pressure measured at the Canada--France--Hawaii Telescope (CFHT) weather station\footnote{The average relative humidity, temperature, and pressure during our observations of PZ Tel were 79.7\%, $3.053^\circ\mathrm{C}$, and \SI{618.7}{mb}. The CFHT weather station archive can be found at \href{http://mkwc.ifa.hawaii.edu/archive/wx/cfht/}{http://mkwc.ifa.hawaii.edu/archive/wx/cfht/}.} at the time of the exposure and the NIRC2 $K_s$ central wavelength of \SI{2.146}{\mu m}.\footnote{\href{http://www2.keck.hawaii.edu/inst/nirc2/filters.html}{http://www2.keck.hawaii.edu/inst/nirc2/filters.html}} For the zenith separation, we project the angular distance between PZ Tel A and B for each individual frame onto its zenith axis determined prior to DAR correction. The average zenith separation for our sequence is $\SI{235}{mas}$ and the average monochromatic DAR is \SI{0.35}{mas}. 

For the chromatic DAR, we use BT-Settl model spectra \citep{allardModelsVerylowmassStars_2012, allardAtmospheresVeryLowmass_2012} to compute the effective wavelength for each source through the $K_s$ filter via the equation
\begin{align}
    \lambda_{\mathrm{eff}} = \frac{\int \lambda I(\lambda) T(\lambda) d\lambda}{\int I(\lambda) T(\lambda) d\lambda},
\end{align}
where $I(\lambda)$ is the model spectrum and $T(\lambda)$ is the filter transmission profile. For the host star, we adopt a $T_{\mathrm{eff}} = \SI{5200}{K}$ and $\log g = \SI{4.0}{dex}$ spectrum based on the effective temperature and surface gravity of PZ Tel A ($T_{\mathrm{eff}} = \SI{5173}{K}$, $\log g = \SI{4.16}{dex}$; \citealt{luckAbundancesLocalRegion_2018}). For PZ Tel B, \citet{maireFirstLightVlt_2016} determined an effective temperature of \SI{2700 \pm 100}{K} and surface gravity $\log g < \SI{4.5}{dex}$. We thus adopt a model spectrum with $T_{\mathrm{eff}} = \SI{2700}{K}$ and $\log g = \SI{4.0}{dex}$ for this purpose. \texttt{refco} is then used to compute refraction constants for both sources, given their $K_s$-band effective wavelengths and the CFHT weather parameters. Finally, we use Equation 11 of \citet{gublerDifferentialAtmosphericRefraction_1998} to compute the chromatic DAR for each frame. The average chromatic DAR across the sequence is \SI{1.54}{mas}. The amount of relative refraction for PZ Tel B is smaller than PZ Tel A. Since PZ Tel B is at a lower zenith angle (higher elevation) in our imaging, chromatic DAR reduces their apparent separation along the zenith axis. To correct our astrometry, we adjust the astrometry measured for each individual frame and increase the separation along the zenith angle by the sum of the two DAR components for that image. Averaged across all frames, this produces a separation of $\rho = 636.1 \pm 1.7 \, \mathrm{mas}$ and position angle of $\theta = 59\fdg33 \pm 0\fdg10$. Correcting for DAR increases $\rho$ by \SI{0.7}{mas} and decreases $\theta$ by 0\fdg16.

The partially transparent coronagraph mask has been found to introduce additional systematic uncertainty on the host-star position \citep[e.g.,][]{konopackyAstrometricMonitoringHr_2016}. \citet{bowlerOrbitDynamicalMass_2018} found that the \SI{600}{mas} mask imparts astrometric uncertainty at the $4{-}\SI{5}{mas}$-level in separation based on the observed astrometric jitter of the brown dwarf companion Gl 758 B. This corresponds to $\pm 0\fdg4{-}0\fdg5$ in position angle at the separation of PZ Tel B. This may impact our astrometry, so we conservatively adopt a noise floor of $\pm \SI{5}{mas}$ in separation and $\pm 0\fdg5$ in position angle. Our final astrometry is then $\rho = \SI{636 \pm 5}{mas}$ and $\theta = 59\fdg3 \pm 0\fdg5$.

\begin{figure*}
    \centering
    \includegraphics[width=0.9\textwidth]{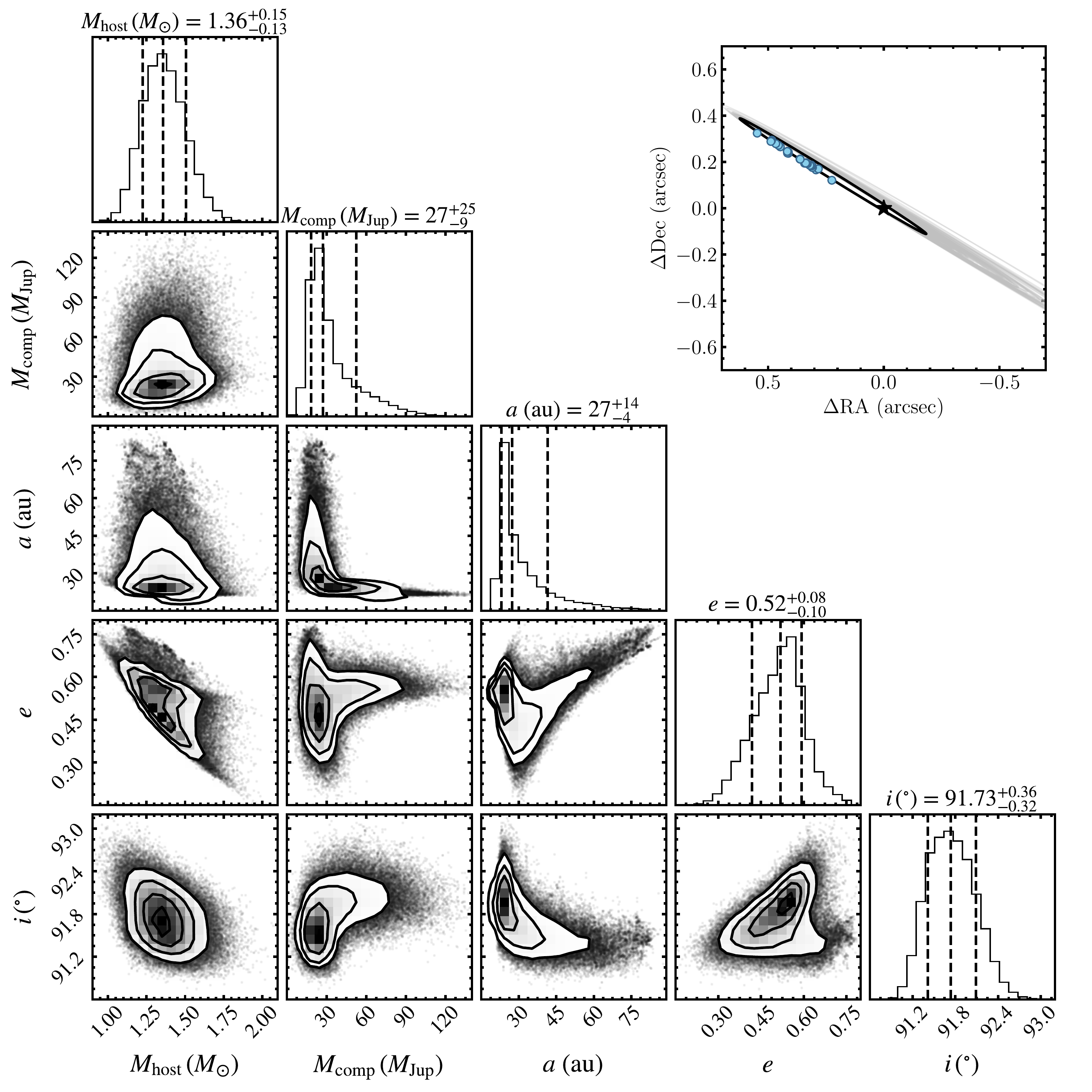}
    \caption{Joint posterior distributions for host-star mass ($M_{\mathrm{host}}$), companion mass ($M_{\mathrm{comp}}$), semi-major axis ($a$), eccentricity ($e$), and inclination ($i$) for the orbit fit of PZ Tel B. Diagonal panels show the marginalized distributions for each parameter. Off-diagonal panels show the covariance between orbital elements. The plot in the upper right shows the sky-projected orbit of PZ Tel B over time. The gray curves are drawn from the MCMC chains, while the black curve highlights the maximum-likelihood orbit. We measure a dynamical mass of $27^{+25}_{-9} \, M_{\mathrm{Jup}}$ for PZ Tel B.
    \label{fig:corner}}
\end{figure*}

\begin{figure*}
    \centering
    \includegraphics[width=0.9\textwidth]{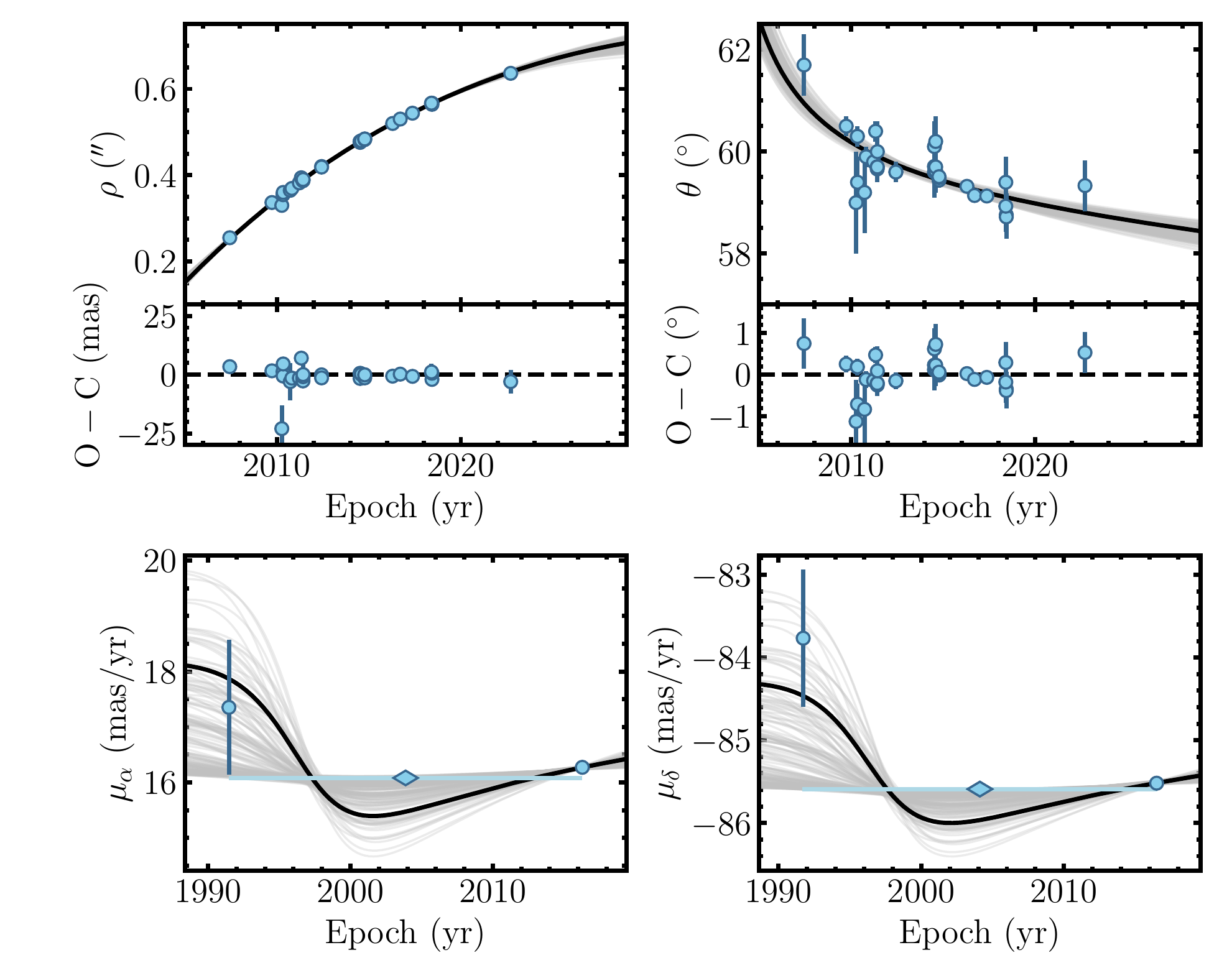}
    \caption{Comparison of relative and absolute astrometry against the PZ Tel B orbit fit. The upper plots show the separation ($\rho$) and position angle ($\theta$) of PZ Tel B over time. The lower plots display the HGCA proper motions. Note that the middle point at ${\approx}2004$ is a joint proper motion that averages the astrometric reflex motion of PZ Tel A over the 25 yr time baseline between Hipparcos and Gaia EDR3. The gray curves are drawn from the MCMC chains, while the black curves highlight the maximum-likelihood orbit.
    \label{fig:orbitfit}}
\end{figure*}

\begin{deluxetable*}{lccc} 
\tablecaption{\label{tab:elements}PZ Tel B Orbit Fit Results}
\tablehead{\colhead{Parameter} & \colhead{Median $\pm 1\sigma$} & \colhead{95.4\% C.I.} & \colhead{Prior}}
\startdata
\multicolumn{4}{c}{Fitted Parameters} \\
\hline
$M_{\mathrm{comp}}$ $(M_\mathrm{Jup})$ & ${27}_{-9}^{+25}$ & (12, 88) & $1/M_{\mathrm{comp}}$ (log-flat)\\
$M_{\mathrm{host}}$ $(M_\mathrm{\odot})$ & ${1.36}_{-0.13}^{+0.15}$ & (1.11, 1.67) & $\SI{1.1 \pm 0.2}{M_\odot}$ (Gaussian)\\
$a$ $(\mathrm{AU})$ & ${27}_{-4}^{+14}$ & (22, 66) & $1/a$ (log-flat)\\
$i$ $(\si{\degree})$ & ${91.73}_{-0.32}^{+0.36}$ & (91.17, 92.43) & $\sin (i)$, $\SI{0}{\degree} < i < \SI{180}{\degree}$\\
$\sqrt{e} \sin{\omega}$ & ${0.2}_{-0.8}^{+0.4}$ & (-0.8, 0.8) & Uniform\\
$\sqrt{e} \cos{\omega}$ & $0.0 \pm 0.6$ & (-0.7, 0.7) & Uniform\\
$\Omega$ $(\si{\degree})$ & ${238.62}_{-0.23}^{+0.21}$\tablenotemark{a} & (238.17, 238.99)\tablenotemark{a} & Uniform\\
$\lambda_{\mathrm{ref}}$ $(\si{\degree})$\tablenotemark{b} & ${110}_{-40}^{+160}$ & (70, 290) & Uniform\\
Parallax $(\si{mas})$ & $21.163 \pm 0.031$ & (21.101, 21.224) & $\SI{21.162 \pm 0.022}{mas}$ (Gaussian)\\
$\mu_\alpha$ ($\si{mas.yr^{-1}}$) & ${16.63}_{-0.12}^{+0.34}$ & (16.41, 17.40) & Uniform\\
$\mu_\delta$ ($\si{mas.yr^{-1}}$) & ${-85.27}_{-0.08}^{+0.21}$ & (-85.42, -84.80) & Uniform\\
RV Jitter $\sigma_{\mathrm{RV}}$ ($\si{m.s^{-1}}$) & ${520}_{-50}^{+60}$ & . . . & $1/\sigma_{\mathrm{RV}}$ (log-flat), $\sigma_{\mathrm{RV}} \in (0, 1000\si{m.s^{-1}}]$\\
\hline
\multicolumn{4}{c}{Derived Parameters} \\
\hline
$P$ (yr) & ${120}_{-30}^{+110}$ & (80, 470) & . . .\\
$e$ & ${0.52}_{-0.10}^{+0.08}$ & (0.32, 0.69) & . . .\\
$\omega$ $(\si{\degree})$ & ${50}_{-30}^{+50}$\tablenotemark{c} & (10, 120)\tablenotemark{c} & . . .\\
$T_0$ $(\mathrm{JD})$ & ${2450500}_{-700}^{+1700}$ & (2449200, 2454000) & . . .\\
$T_0$ (yr) & ${1997.2}_{-1.8}^{+4.7}$ & (1993.7, 2006.7) & . . .\\
$q$ $(=M_{\mathrm{comp}}/M_{\mathrm{host}})$ & ${0.019}_{-0.006}^{+0.018}$ & (0.009, 0.062) & . . .
\enddata
\tablenotetext{a}{The posterior distribution for $\Omega$ consists of two distinct peaks separated by \SI{180}{\degree}. The values shown in the table correspond to the higher peak. The other peak is located at ${58.61}_{-0.22}^{+0.20}{}^\circ$ with a 95.4\% confidence interval of (58.17, 58.97).}\tablenotetext{b}{Mean longitude at the reference epoch of 2010.0.}
\tablenotetext{c}{The posterior distribution for $\omega$ is bimodal, with two peaks with similar shapes separated by \SI{180}{\degree}. The values shown in the table correspond to the slightly higher peak. The other peak is located at ${220}_{-20}^{+50}{}^\circ$ with a 95.4\% confidence interval of (190, 290).}

\end{deluxetable*}

\section{3D Orbit Fit}
To measure the dynamical mass of PZ Tel B, we perform a joint orbit fit of all available relative astrometry, radial velocities (RVs), and the absolute astrometry from Hipparcos and Gaia EDR3 in the HGCA. The published relative astrometry of this system comprises 33 measurements from June 2007 to June 2018. This astrometry was previously assembled in \citet{bowlerPopulationlevelEccentricityDistributions_2020}, which included epochs from \citet{billerGeminiNiciPlanetfinding_2010}, \citet{mugrauerNewObservationsPz_2012}, \citet{ginskiAstrometricFollowupObservations_2014}, \citet{beustOrbitalFittingImaged_2016}, and \citet{maireFirstLightVlt_2016}. We supplement the \citet{bowlerPopulationlevelEccentricityDistributions_2020} compilation with three additional epochs from 2016 taken in the SHINE survey \citep{langloisSphereInfraredSurvey_2021}, two measurements from May 2018 reported in \citet{mussobarcucciDetectionEmissionPz_2019}, and two measurements from June 2018 reported in \citet{stolkerMiraclesAtmosphericCharacterization_2020}. We also include our new September 2022 Keck/NIRC2 astrometry in the orbit fit for a total of 34 data points.

Precise radial velocity measurements of PZ Tel A are obtained from the \texttt{HARPS-RVBANK} archive \citep{trifonovPublicHarpsRadial_2020}. They total 42 measurements taken between April 2009 and October 2017 with the High Accuracy Radial velocity Planet Searcher \citep[HARPS;][]{pepeHarpsEsoComing_2002,mayorSettingNewStandards_2003}, a high-resolution ($R\approx 115,000$) optical spectrograph mounted on the \SI{3.6}{m} ESO telescope at La Silla Observatory. The RVs have an rms value of \SI{513}{m.s^{-1}} and a median uncertainty on each individual measurement of $\SI{4.2}{m.s^{-1}}$. The dominant sources of instrumental and astrophysical noise are likely stellar activity due to the star's youth and broadening from its fast rotation ($v\sin i = \SI{73 \pm 5}{km.s^{-1}}$; \citealt{jenkinsBenchmarkCoolCompanions_2012}). We use the measurements in the catalog under the ``DRVmlcnzp'' column, which incorporates corrections for nightly zero-point variations, intra-night RV drift, and a discontinuity in the absolute RV associated with the May 2015 upgrade of the HARPS fibers \citep{locurtoHarpsGetsNew_2015}.

Our joint orbit fit is carried out with the \texttt{orvara} orbit fitting package \citep{brandtOrvaraEfficientCode_2021}, which uses the parallel-tempered Markov chain Monte Carlo (PT-MCMC) ensemble sampler in \texttt{emcee} \citep{foreman-mackeyEmceeMcmcHammer_2013} to sample the orbit element posterior parameter space. \texttt{orvara} directly fits the following quantities: the host star mass ($M_{\mathrm{host}}$), the companion mass ($M_{\mathrm{comp}}$), semi-major axis ($a$), inclination ($i$), longitude of ascending node ($\Omega$), the longitude at the reference epoch of 2010.0 ($\lambda_{\mathrm{ref}}$), and an RV jitter term ($\sigma_{\mathrm{RV}}$). Eccentricity ($e$) and argument of periastron ($\omega$) are parameterized as $\sqrt{e} \sin{\omega}$ and $\sqrt{e} \cos{\omega}$ in the fit to avoid the Lucy-Sweeney bias against circular orbits \citep{lucySpectroscopicBinariesCircular_1971}. \texttt{orvara} analytically marginalizes over the parallax, barycenter proper motion, and instrumental RV zero-points to increase computational efficiency.

We adopt uninformative priors for all orbital elements and the companion mass. The host-star mass prior is assigned a Gaussian distribution of \SI{1.1 \pm 0.2}{M_\odot} based on typical values for the mass of PZ Tel A in the literature (e.g., $1.25^{+0.05}_{-0.20} \, \mathrm{M_{\odot}}$ from \citealt{dantonaNewPreMainsequence_1994}; $1.02 \pm 0.04 \, \mathrm{M_{\odot}}$ from \citealt{allendeprietoFundamentalParametersNearby_1999}, $1.2 \pm 0.1 \, \mathrm{M_{\odot}}$ from \citealt{tetzlaffCatalogueYoungRunaway_2011}; \SI{1.13 \pm 0.03}{M_\odot} from \citealt{jenkinsBenchmarkCoolCompanions_2012}; \SI{1.14}{M_{\odot}} from \citealt{zuniga-fernandezSearchAssociationsContaining_2021}). The Gaia EDR3 value of \SI{21.162 \pm 0.022}{mas} \citep{gaiacollaborationGaiaEarlyData_2021} is used for the parallax prior. Our priors for all fitted orbit elements and physical parameters are shown in Table \ref{tab:elements}.

We use a total of 100 walkers, 20 temperatures, and $10^6$ total steps to sample the parameter space. The posterior distribution for the parallel-tempered sampling algorithm is described by the coldest-temperature chain \citep[see e.g.,][]{vousdenDynamicTemperatureSelection_2016}. We discard the first 50\% (5000 steps of each walker) as burn-in. Table \ref{tab:elements} shows the marginalized parameter posteriors from the orbit fit. Figure \ref{fig:corner} displays the posterior distributions for select orbital elements. A comparison of the relative astrometry of PZ Tel B and HGCA proper motions to a swarm of orbits drawn from the orbit fit is shown in Figure \ref{fig:orbitfit}. We measure a dynamical mass of $27^{+25}_{-9} \, M_{\mathrm{Jup}}$, semi-major axis of $27^{+14}_{-4} \, \mathrm{au}$, inclination of $91.73^{+0.36}_{-0.32} {}^\circ$, and eccentricity of $0.52^{+0.08}_{-0.10}$. The orbital period is $120^{+110}_{-30}\, \mathrm{yr}$ and the time of periastron is $T_0 = 1997.2^{+4.7}_{-1.8}$. 

There is covariance between the companion mass and semi-major axis posteriors: the high-mass tail in companion mass occurs when semi-major axis is small, while the tail to large semi-major axes is restricted to low values of the dynamical mass. This is caused by the similarity of the Hipparcos-Gaia joint proper motion to the Gaia EDR3 proper motion in combination with the limited amount of position angle change over the relative astrometry baseline (see Figure \ref{fig:orbitfit}). While the proper motions are significantly different at the $4\sigma$-level, there remains a degeneracy between large-semi-major-axis solutions, where the proper motion change is small between Hipparcos and Gaia, and small-semi-major-axis solutions, in which the proper motion varies more substantially over the time baseline, but averages to the same value for the joint proper motion. Typically, this degeneracy would be broken by the relative astrometry. However, the small amount of position angle change limits the ability to distinguish these two solutions for PZ Tel B. The large-semi-major-axis orbits from the fit are only permitted if the companion mass is lower ($\lesssim 30 \, M_\mathrm{Jup}$), since higher masses and longer orbital periods would cause more significant changes in the absolute astrometry that would not average to the value of the observed joint proper motion.

The orbit fit prefers higher values of the primary mass ($1.36^{+0.15}_{-0.13} \, \mathrm{M_\odot}$) than our prior of \SI{1.1 \pm 0.2}{M_\odot}. To assess the impact of this prior on the resultant orbit elements, we conduct orbit fits with a narrow (\SI{1.1 \pm 0.1}{M_\odot}) and a wide (\SI{1.1 \pm 0.5}{M_\odot}) prior on the host-star mass. For these fits, we use the same number of walkers, temperatures, and steps as our adopted orbit fit, and again discard the first 50\% of each walker as burn-in. Both priors yield consistent orbit elemenents within the uncertainties. The narrow prior produces $M_{\mathrm{comp}} = 25_{-8}^{+26} \, M_{\mathrm{Jup}}$, $a = 28^{+16}_{-4} \, \mathrm{au}$, $e = 0.58^{+0.06}_{-0.07}$, and $i = 91.89^{+0.35}_{-0.34} {}^\circ$. The wide prior yields $M_{\mathrm{comp}} = 34_{-12}^{+26} \, M_{\mathrm{Jup}}$, $a = 25_{-3}^{+8} \, \mathrm{au}$, $e = 0.42_{-0.16}^{+0.10}$, and $i = 91.55_{-0.28}^{+0.32} {}^\circ$. We thus find that the choice of host-star mass prior has a small impact on our dynamical mass and orbit fit.

\begin{figure*}
    \centering
    \includegraphics[width=\textwidth]{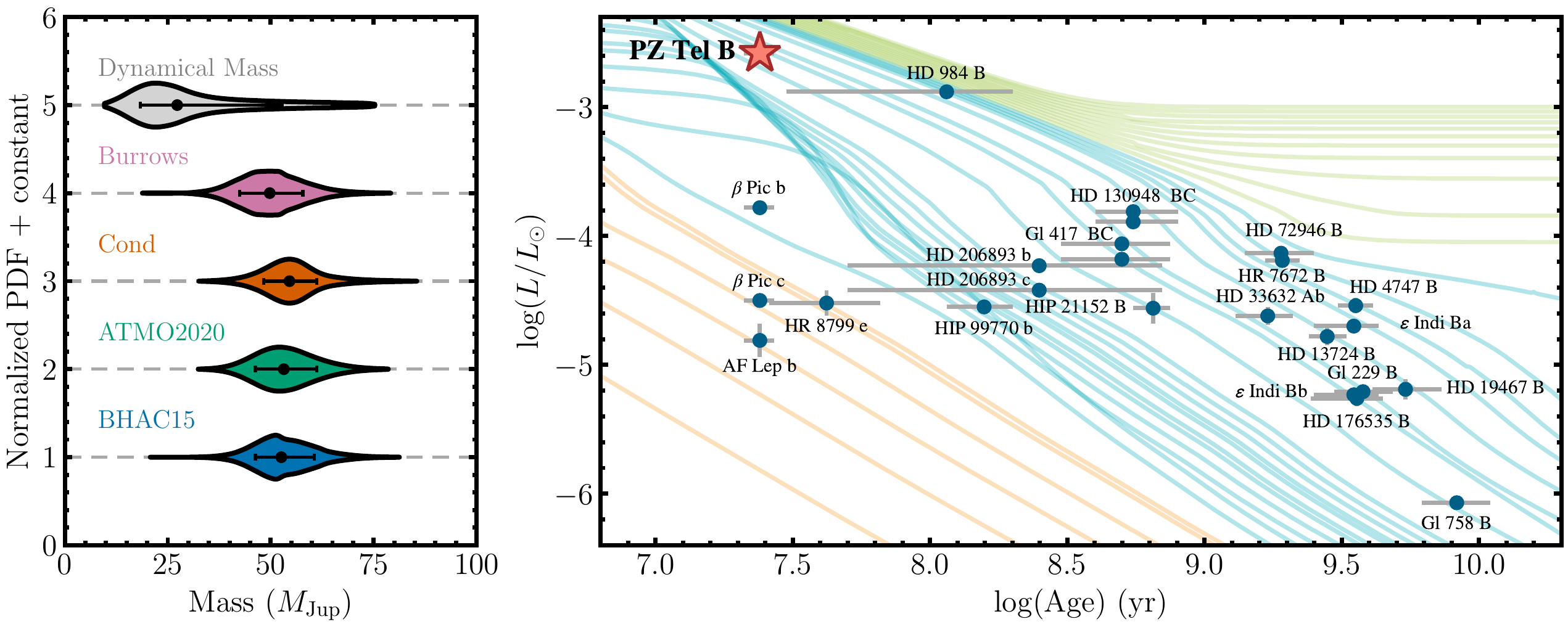}
    \caption{\emph{Left:} Comparison of the dynamical mass of PZ Tel B with the predicted masses from four hot-start evolutionary models. The median and 68.3\% confidence interval are highlighted for each distribution. The inferred masses are consistent with the dynamical mass to within $1\sigma$. \emph{Right:} Dynamical masses with well-constrained ages and luminosities. The background models show the luminosity evolution of planets (orange), brown dwarfs (blue) and low-mass stars (green) of a given mass from \citet{burrowsNongrayTheoryExtrasolar_1997}. PZ Tel B (highlighted in red) is among the few systems with a precise dynamical mass and a well-constrained age under $200\, \mathrm{Myr}$. The luminosities and ages for the majority of systems are tabulated in Table 1 of \citet{fransonDynamicalMassYoung_2022}. We also include HD 206893 bc \citep{hinkleyDirectDiscoveryInner_2022}, HIP 21152 B \citep{bonavitaResultsCopainsPilot_2022,kuzuharaDirectImagingDiscovery_2022,fransonAstrometricAccelerationsDynamical_2023}, $\varepsilon$ Indi Bab \citep{chenPreciseDynamicalMasses_2022}, HIP 99770 b \citep{currieDirectImagingAstrometric_2022}, HD 176535 B \citep{liSurveyingNearbyBrown_2023}, and AF Lep b \citep{fransonAstrometricAccelerationsDynamical_2023a,mesaAfLepLowest_2023,derosaDirectImagingDiscovery_2023}. The age for HD 206893 bc is from \citet{delormeIndepthStudyModerately_2017} and luminosities of $\varepsilon$ Indi Bab are from \citet{kingMathsfVarepsilonIndi_2010}. The luminosity from \citet{fransonAstrometricAccelerationsDynamical_2023a} is adopted for AF Lep b.
    \label{fig:model_comp}}
\end{figure*}

\section{Discussion}

\subsection{Mass of PZ Tel B}
The dynamical mass of $27^{+25}_{-9} \, M_{\mathrm{Jup}}$ is right-skewed, with a tail that extends to significantly higher masses. The hydrogen-burning limit (HBL), defined as the mass at which 50\% of an object's energy is generated via hydrogen fusion, marks the boundary between brown dwarfs and low-mass stars \citep{reidNewLightDark_2005}. The precise value can vary depending on assumptions about the equation of state, rotation, composition, and atmospheric properties \citep{burrowsNongrayTheoryExtrasolar_1997}, but generally ranges from 70--$80 \, M_\mathrm{Jup}$ \citep[e.g.,][]{saumonEvolutionDwarfsColor_2008,scuflaireClesCodeLiegeois_2008,baraffeNewEvolutionaryModels_2015,dupuyIndividualDynamicalMasses_2017}. Adpoting $75 \, M_\mathrm{Jup}$ as the substellar boundary, 95.0\% of the dynamical mass posterior is below this value. At a lower HBL threshold $70 M_\mathrm{Jup}$, 93.3\% of the dynamical mass is below the value. Similar to the HBL, the deuterium-burning limit (DBL), which traditionally demarcates planets and brown dwarfs \citep{bossNomenclatureBrownDwarfs_2003,bossWorkingGroupExtrasolar_2007}, can vary based on a given objects' helium abundance, initial deuterium abundance, and metallicity, but generally is ${\approx} 13 \, M_{\mathrm{Jup}}$ \citep{spiegelDeuteriumburningMassLimit_2011,molliereDeuteriumBurningObjects_2012}. A total of 96.5\% of the dynamical mass posterior falls above $13 \, M_\mathrm{Jup}$, while 91.5\% lies between $13 \, M_\mathrm{Jup}$ and $75 \, M_\mathrm{Jup}$. We thus find that the dynamical mass of PZ Tel B is almost surely in the brown dwarf regime. 

\subsection{Orbit of PZ Tel B}
Our orbit fit broadly recovers the high eccentricity long associated with PZ Tel B \citep[e.g.,][]{billerGeminiNiciPlanetfinding_2010,mugrauerNewObservationsPz_2012}, although our eccentricity posterior of $0.52^{+0.08}_{-0.10}$ is significantly lower than many previous determinations. \citet{bowlerPopulationlevelEccentricityDistributions_2020} and \citet{beustOrbitalFittingImaged_2016} found a higher eccentricity values of $0.89^{+0.10}_{-0.05}$ and $e > 0.91$, resepectively, which were consistent with previous bounds of $e > 0.6$ from \citet{billerGeminiNiciPlanetfinding_2010}, $0.66 \leq e \leq 0.99$ from \citet{ginskiAstrometricFollowupObservations_2014}, and $e \gtrsim 0.62$ from \citet{maireFirstLightVlt_2016}. Our eccentricity is most similar to \citet{mussobarcucciDetectionEmissionPz_2019}, who found the eccentricity distribution to be multi-modal with a larger peak at $e \sim 0.55$ and smaller peak at $e \sim 1$. Note that the orbit fit from \citet{mussobarcucciDetectionEmissionPz_2019} has the longest baseline of relative astrometry (2007--2018) among the fits in the literature. The semi-major axis of $27^{+14}_{-4} \, \mathrm{au}$ that we find is compatibile with previous orbit fits (e.g., $a \approx \SI{31.3}{au}$, \citealt{mussobarcucciDetectionEmissionPz_2019}; $a = 24.8^{+5.3}_{-5.5} \, \mathrm{au}$, \citealt{bowlerPopulationlevelEccentricityDistributions_2020}; $a \gtrsim \SI{24.5}{au}$, \citealt{maireFirstLightVlt_2016}).

The inclination from our orbit fit of $91.73^{+0.36}_{-0.32} {}^\circ$ is consistent with the results of previous fits which identified the companion as having an edge-on orbit (e.g., $i \sim 91.6^\circ$, \citealt{mussobarcucciDetectionEmissionPz_2019}; $i = 93.4^{+1.2}_{-1.7} {}^\circ$, \citealt{bowlerPopulationlevelEccentricityDistributions_2020}; $\SI{91}{\degree} < i < \SI{96.1}{\degree}$, \citealt{maireFirstLightVlt_2016}). \citet{bowlerRotationPeriodsInclinations_2023} determined the inclination of PZ Tel A to be $i_* = 78.9^{+11.0}_{-4.7} {}^\circ$. This is compatibile with alignment between the host star's rotational axis and PZ Tel B's orbit, given our measurement for the companion's orbital inclination. Note, though, that misalignment remains a possibility. To conclusively determine the mutual alignment between the orbit of PZ Tel B and the rotation of its host star, one must determine the orientation of PZ Tel A's rotation axis on the sky-plane \citep[see e.g.,][]{krausSpinorbitAlignmentBeta_2020}. 

\subsection{Comparison with Evolutionary Model Predictions}
Here, we compare the dynamical mass of PZ Tel B with the predicted masses from substellar evolutionary models that cover the age and luminosity of the companion. We consider four hot-start evolutionary models: the \citet{burrowsNongrayTheoryExtrasolar_1997} grid, \texttt{Cond} \citep{baraffeEvolutionaryModelsCool_2003}, \texttt{ATMO-2020} \citep{phillipsNewSetAtmosphere_2020}, and \texttt{BHAC-15} \citep{baraffeNewEvolutionaryModels_2015}.\footnote{Note that the \citet{saumonEvolutionDwarfsColor_2008} models do not extend to objects as luminous as PZ Tel B.}

Our approach is to compute inferred masses for each evolutionary model given the age and luminosity of the companion. For the companion age, we adopt the age of the $\beta$ Pic moving group of \SI{24 \pm 3}{Myr} from \citet{bellSelfconsistentAbsoluteIsochronal_2015}. We compute the bolometric luminosity by applying a $J$-band bolometric correction of $\mathrm{BC}_{J_{\mathrm{2MASS}}} = \SI{2.06 \pm 0.05}{mag}$\footnote{\citet{herczegEmpiricalIsochronesLow_2015} give a general uncertainty range on their bolometric corrections of $0.02{-}0.05 \, \mathrm{mag}$. For this analysis, we conservatively adopt the upper value for the uncertainty.} from \citet{herczegEmpiricalIsochronesLow_2015} based on the $\mathrm{M7} \pm 1$ spectral type of PZ Tel B \citep{maireFirstLightVlt_2016}. \citet{liuHawaiiInfraredParallax_2016} measured $J$-band photometry in the Mauna-Kea Observatories (MKO) system of \SI{12.47 \pm 0.20}{mag}, which corresponds to $J_{\mathrm{2MASS}} = \SI{12.52 \pm 0.20}{mag}$. With the Gaia DR3 parallax of \SI{21.1621 \pm 0.0223}{mas} \citep{gaiacollaborationGaiaDataRelease_2022}, the absolute magnitude of PZ Tel B $M_{J_{\mathrm{2MASS}}} = \SI{9.15 \pm 0.20}{mag}$, which yields a bolometric luminosity of $\SI{-2.59 \pm 0.08}{dex}$. This is consistent with the bolometric luminosity of $\SI{-2.51 \pm 0.10}{dex}$ determined in \citet{maireFirstLightVlt_2016} using $K_s$ photometry and the $K$-band bolometric correction from \citet{liuDiscoveryHighlyUnequalmass_2010}. We adopt our new bolometric luminosity value for the model comparison.

To generate inferred masses for each evolutionary model, we employ a Monte Carlo approach by randomly sampling from the bolometric luminosity and age distributions. The model grid is then linearly interpolated to determine the companion mass that corresponds to the specific age and luminosity value. This process is then repeated $10^6$ times to build up a model-inferred mass distribution. Figure \ref{fig:model_comp} shows the inferred masses for the four grids we consider alongside the dynamical mass of PZ Tel B. To quantify the level of agreement between the dynamical mass and inferred masses, we follow \citet{fransonAstrometricAccelerationsDynamical_2023} in determining $P(M_{\mathrm{Inferred}} > M_{\mathrm{Dynamical}})$, the probability that the inferred mass distribution function is greater than the dynamical mass distribution. Probabilities of $\approx 50\%$ signify that the two distributions are consistent with one another, while values near 0\% or 100\% correspond to the inferred mass being lower or higher than the dynamical mass, respectively. $P(M_{\mathrm{Inferred}} > M_{\mathrm{Dynamical}})$ can also be converted to a one-sided Gaussian-equivalent standard deviation, $\sigma$, via the equation 
\begin{align}
    \sigma = \sqrt{2} \, \mathrm{erf}^{-1}(1 - 2 \, P(M_{\mathrm{Inferred}} > M_{\mathrm{Dynamical}})).
\end{align}
For \citet{burrowsNongrayTheoryExtrasolar_1997}, $P(M_{\mathrm{Inferred}} > M_{\mathrm{Dynamical}}) = 82.0\%$, which corresponds to $+0.9\sigma$. For \texttt{Cond}, $P(M_{\mathrm{Inferred}} > M_{\mathrm{Dynamical}}) = 85.4\%$ ($+1.1\sigma$). For \texttt{BHAC-15}, $P(M_{\mathrm{Inferred}} > M_{\mathrm{Dynamical}}) = 84.4\%$ ($+1.0\sigma$). Finally, for \texttt{ATMO-2020}, $P(M_{\mathrm{Inferred}} > M_{\mathrm{Dynamical}}) = 84.6\%$ ($+1.0\sigma$). Overall, we find that all models are consistent with our dynamical mass to within about $1 \sigma$.

\begin{figure*}
    \centering
    \includegraphics[width=\linewidth]{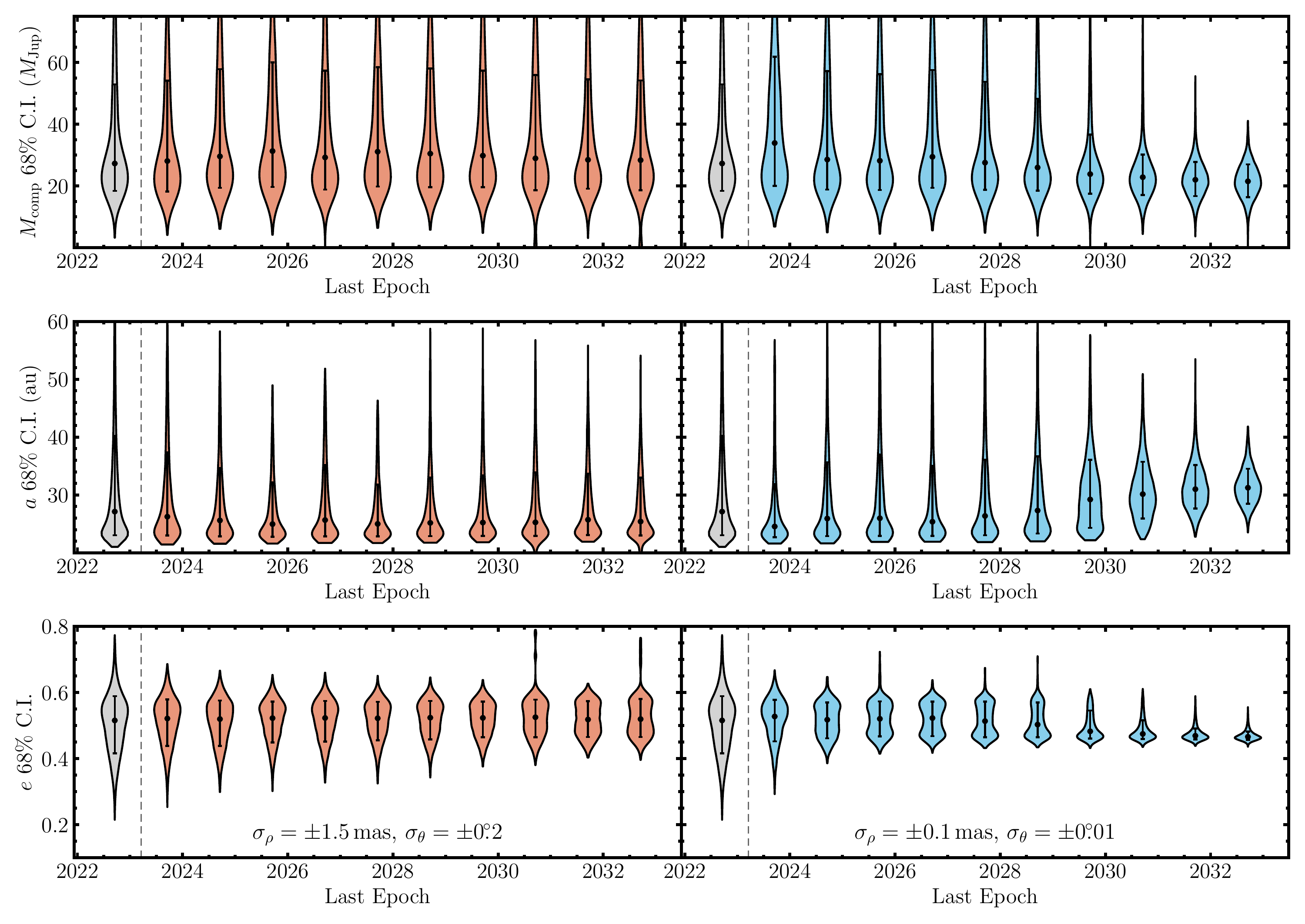}
    \caption{Companion mass (top), semi-major axis (middle), and eccentricity (bottom) posteriors from orbit fits with additional epochs of mock relative astrometry. The left panels (red distributions) show runs with additional epochs of $1.5\, \mathrm{mas}$-precision astrometry (e.g., SPHERE, SCExAO, GPI), while the right panels (blue distributions) display mock runs with additional epochs of $0.1\, \mathrm{mas}$-precision astrometry (e.g., GRAVITY). The mock astrometry is added with a 1-year cadence; each distribution shows an orbit fit with annual epochs from 2023.77 to its last epoch. The posteriors from our orbit fit with no mock data are shown in gray. The median and 68.3\% confidence interval are highlighted for each distribution.
    \label{fig:new_astrometry}}
\end{figure*}

\subsection{Improving the Dynamical Mass Precision}
Finally, we explore the prospects for improving the precision of the dynamical mass of PZ Tel B with additional epochs of astrometric orbit monitoring. Our approach is to conduct orbit fits with mock astrometry at typical astrometric precisions of extreme AO (e.g., SPHERE, \citealt{beuzitSphereExoplanetImager_2019}; SCExAO, \citealt{jovanovicSubaruCoronagraphicExtreme_2015}, GPI, \citealt{macintoshFirstLightGemini_2014}) and GRAVITY \citep{gravitycollaborationFirstLightGravity_2017} and examine how the dynamical mass posterior changes over time. For the extreme AO case, we adopt an uncertainty of $\pm \SI{1.5}{mas}$ in separation and $\pm 0\fdg2$ in position angle \citep[e.g.,][]{maireFirstLightVlt_2016,johnson-grohIntegralFieldSpectroscopy_2017}. For GRAVITY astrometry, a typical positional uncertainty of $\pm \SI{0.1}{mas}$ \citep[e.g.,][]{nowakDirectConfirmationRadialvelocity_2020,hinkleyDirectDiscoveryInner_2022} corresponds to an uncertainty of $\pm \SI{0.1}{mas}$ in separation and $\pm 0\fdg01$ in position angle. 

Starting at $t = 2023.710$ (one year after our new Keck/NIRC2 epoch), we add new epochs of relative astrometry at a cadence of one mock observation per year. The separation and position angle for a given mock epoch are taken from the highest-likelihood orbit from our joint orbit fit with $M_{\mathrm{comp}}$, $a$, and $e$ all within $1\,M_{\mathrm{Jup}}$, $1 \, \mathrm{au}$, and $0.05$ of their mean values in their respective posterior distributions. For each epoch, a new joint orbit fit with \texttt{orvara} is performed incorporating all of the previous epochs of mock relative astrometry alongside the true astrometry, radial velocities, and HGCA proper motions of the system. For these orbit fits, we use 20 temperatures, 100 walkers, and $10^5$ total steps. We discard the first 50\% of each chain as burn-in. 

Figure \ref{fig:new_astrometry} shows the dynamical mass, semi-major axis, and eccentricity posteriors from our mock orbit fits. We find that at extreme-AO-level precisions, the constraints on companion mass, semi-major axis, and eccentricity are not meaningfully improved with additional astrometry over this ten-year timeframe. At GRAVITY-level precisions, these constraints can be substantially improved with continued astrometric monitoring. After ${\approx}8$ years of annual GRAVITY measurements, this exercise predicts that the dynamical mass would be constrained to the $\pm 5 \, M_\mathrm{Jup}$ level, the semi-major axis would be constrained to the $\pm 5 \, \mathrm{au}$ level, and the eccentricity would be constrained to $\pm 0.03$. The 68\% credible interval for the companion mass decreases from $34 \, M_\mathrm{Jup}$ to $11 \, M_\mathrm{Jup}$ between our orbit fit with no mock data and an orbit fit with ten epochs of additional GRAVITY astrometry. The credible intervals for the semi-major axis and eccentricity decrease from $18 \, \mathrm{au}$ to $6 \, \mathrm{au}$ and $0.18$ to 0.03, respectively.

There are several aspects of the PZ Tel system that create challenges for improving the mass constraint. The companion is currently approaching apastron, where the separation changes the least year-over-year. Furthermore, the nearly edge-on orbit means that the annual change in position angle is small. The host star's fast projected rotation ($v\sin i = \SI{73 \pm 5}{km.s^{-1}}$; \citealt{jenkinsBenchmarkCoolCompanions_2012}) and youth make the detection of an RV acceleration challenging. Despite these challenges, astrometric monitoring with GRAVITY is capable of significantly increasing the dynamical mass precision in the near future. Additional avenues for improving the mass constraint include measuring the RV of PZ Tel B with the Keck Planet Imager and Characterizer \citep[e.g.,][]{wangDetectionBulkProperties_2021} and future Gaia monitoring of the host star.

\section{Summary}
In this work, we measured the dynamical mass and orbit of the young brown dwarf companion PZ Tel B by combining Hipparcos-Gaia proper motions, relative astrometry including new Keck/NIRC2 imaging, and archival HARPS RVs. Our joint orbit fit produces a dynamical mass of $27^{+25}_{-9} \, M_\mathrm{Jup}$, semi-major axis of $27^{+14}_{-4} \, \mathrm{au}$, eccentricity of $0.52^{+0.08}_{-0.10}$, and inclination of $91.73^{+0.36}_{-0.32} {}^\circ$. The model-independent mass is consistent within $1.1\sigma$ of predicted masses from hot-start evolutionary models given the companion's age and luminosity. The eccentricity from our fit is significantly lower than many previous orbit fits, which generally identified eccentricities above 0.6. We also examined the impact of additional epochs of relative astrometry on the precision of the dynamical mass. High-precision astrometric monitoring with GRAVITY is capable of improving the mass constraint by a factor of three over ten epochs. PZ Tel B joins $\beta$ Pic b, $\beta$ Pic c, and AF Lep b as the only substellar companions in young associations with dynamical masses.

\section{acknowledgements}
K.F. acknowledges support from the National Science Foundation Graduate Research Fellowship Program under Grant No. DGE-1610403. B.P.B. acknowledges support from the National Science Foundation grant AST-1909209, NASA Exoplanet Research Program grant 20-XRP20$\_$2-0119, and the Alfred P. Sloan Foundation. This work was supported by a NASA Keck PI Data Award, administered by the NASA Exoplanet Science Institute.

This work has made use of data fram the European Space Agency (ESA) space mission Gaia. Gaia data are being processed by the Gaia Data Processing and Analysis Consortium (DPAC). Funding for the DPAC is provided by national institutions, in particular the institutions participating in the Gaia MultiLateral Agreement (MLA). The Gaia mission website is \href{https://www.cosmos.esa.int/gaia}{https://www.cosmos.esa.int/gaia}. The Gaia archive website is \href{https://archives.esac.esa.int/gaia}{https://archives.esac.esa.int/gaia}. This research has made use of the SIMBAD database and the VizieR catalogue access tool, CDS, Strasbourg, France. The Starlink software \citep{currieStarlinkSoftware2013_2014} is currently supported by the East Asian Observatory.

The authors wish to recognize and acknowledge the very significant cultural role and reverence that the summit of Maunakea has always had within the indigenous Hawaiian community. We are most fortunate to have the opportunity to conduct observations from this mountain.

\facilities{Keck:II (NIRC2)}
\software{\texttt{VIP} \citep{gomezgonzalezVipVortexImage_2017}, \texttt{orvara} \citep{brandtOrvaraEfficientCode_2021}, \texttt{ccdproc} \citep{craigAstropyCcdprocV1_2017}, \texttt{photutils} \citep{bradleyAstropyPhotutilsV0_2019}, \texttt{astropy} \citep{astropycollaborationAstropyCommunityPython_2013,astropycollaborationAstropyProjectBuilding_2018}, \texttt{pandas} \citep{mckinneyDataStructuresStatistical_2010}, \texttt{matplotlib} \citep{hunterMatplotlib2dGraphics_2007}, \texttt{numpy} \citep{harrisArrayProgrammingNumpy_2020}, \texttt{scipy} \citep{virtanenScipyFundamentalAlgorithms_2020}, \texttt{emcee} \citep{foreman-mackeyEmceeMcmcHammer_2013}, \texttt{corner} \citep{foreman-mackeyCornerPyScatterplot_2016}, Starlink \citep{currieStarlinkSoftware2013_2014}, AST \citep{berryAstLibraryModelling_2016}}

\bibliography{references}{}
\bibliographystyle{aasjournal}
\newpage

\end{document}